\documentclass[lnicst]{svmultln}
\usepackage{makeidx}  % allows for indexgeneration
\usepackage{amsmath}
\usepackage{amssymb}
\usepackage{alltt}
\usepackage{epsfig}
\usepackage{listings}
\usepackage{subfig}
\usepackage{float}
\usepackage{wrapfig}
\usepackage{stmaryrd}
\usepackage{enumerate}
\usepackage{url}

\usepackage{textcomp} 

\DeclareMathOperator{\dom}{\mathrm{dom}} 
\DeclareMathOperator{\cod}{\mathrm{cod}} 

\newcommand{\horizfig}[3][]{%
  \begin{minipage}{#2}\centering\subfloat[#1]{#3}\end{minipage}}

\newcommand{\scsem}[1]{{\left\lceil #1 \right\rceil_{DS}}} 
\newcommand{\scsemp}[2]{{\left\lceil #1 \right\rceil_{#2}}} 

\newcommand{\rsem}[1]{{\llceil #1 \rrceil_{DS,\rho}}} 
\newcommand{\rsemp}[2]{{\llceil #1 \rrceil_{#2}}} 

\newcommand{\sigsem}[1]{{\llfloor #1 \rrfloor_{DS,\rho}}} 
\newcommand{\sigsemp}[2]{{\llfloor #1 \rrfloor_{#2}}} 

\newcommand{\pssemp}[2]{{\left\| #1 \right\|_{#2}}}

\newcommand{\targetdef}{target definition} 
\newcommand{\syscomp}{system component} 
\newcommand{\sysinst}{software component identifier}
\newcommand{\softcomp}{software component} 
\newcommand{\relt}{relation} 
 
\newcommand{\propt}{property}
\newcommand{\propts}{properties}
\newcommand{\Ds}{data source}
\newcommand{\ds}{DS}
\newcommand{\mandom}{managed domain}

\newcommand{\product}{\mathtt{product}}
\newcommand{\vendor}{\mathtt{vendor}}
\newcommand{\release}{\mathtt{release}}
\newcommand{\supspec}{\mathtt{sup\_spec}}
\newcommand{\reqspec}{\mathtt{req\_spec}}
\newcommand{\ipadd}{\mathtt{ip\_jmx}}
\newcommand{\port}{\mathtt{port\_jmx}}
\newcommand{\croot}{\mathtt{ctx\_root}}
\newcommand{\uncpath}{\mathtt{unc\_path}}
\newcommand{\deplin}{\mathtt{depl\_in}}
\newcommand{\commwith}{\mathtt{comm\_with}}
\newcommand{\compof}{\mathtt{comp\_of}}
\newcommand{\inset}{\mathtt{instr\_set}}
\newcommand{\SC}{\mathit{SC}}
\newcommand{\OD}{\mathit{OD}}
\newcommand{\TD}{\mathit{TD}}
\newcommand{\SCS}{\mathit{SCS}}
\newcommand{\RS}{\mathit{RS}}
\newcommand{\CD}{\mathit{CD}}
\newcommand{\SI}{\mathit{SI}}
\newcommand{\SIS}{\mathit{SIS}}
\newcommand{\CL}{\mathit{K}}
\newcommand{\PS}{\mathit{PS}}
\newcommand{\ST}{\mathit{ST}}
\newcommand{\TM}{\mathit{TM}}
\newcommand{\CR}{\mathit{CR}}

\makeatletter
\def\old@comma{,}
\catcode`\,=13
\def,{%
  \ifmmode% 
     \old@comma\discretionary{}{}{}%
  \else%
    \old@comma%
  \fi%
}
\makeatother

\begin{document}
\mainmatter              % start of the contribution
\title{Detection of Configuration Vulnerabilities in Distributed (Web) Environments \thanks{This work was partially
    supported by the FP7-ICT-2009.1.4 Project PoSecCo (no.~257129,
    \protect\url{www.posecco.eu})}}

\titlerunning{Detection of configuration vulnerabilities}  % abbreviated title (for running head)
%                                     also used for the TOC unless
%                                     \toctitle is used

\author{Matteo Maria Casalino \and Michele Mangili \and Henrik Plate \and Serena
  Elisa Ponta}

\institute{SAP Research Sophia-Antipolis, 805 Avenue Dr M. Donat,
  06250 Mougins, France \email{{ matteo.maria.casalino,
     henrik.plate, serena.ponta}@sap.com}}

\authorrunning{M. M. Casalino \and M. Mangili \and H. Plate \and S. E. Ponta}   % abbreviated author list (for running head)
\maketitle              % typeset the title of the contribution

\begin{abstract}        % give a summary of your paper
Many tools and libraries are readily available to build and operate
distributed Web applications. While the setup of operational
environments is comparatively easy, practice shows that their
continuous secure operation is more difficult to achieve, many times
resulting in vulnerable systems exposed to the Internet. Authenticated
vulnerability scanners and validation tools represent a means to
detect security vulnerabilities caused by missing patches or
misconfiguration, but current approaches center much around the
concepts of hosts and operating systems. This paper presents a
language and an approach for the declarative specification and
execution of machine-readable security checks for sets of more
fine-granular system components depending on each other in a
distributed environment. Such a language, building on existing
standards, fosters the creation and sharing of security content among
security stakeholders.  Our approach is exemplified by vulnerabilities
of and corresponding checks for Open Source Software commonly used in
today's Internet applications.

\end{abstract}
%%%%%%%%%%%%%%%%%%%%%%%%%%%%%%%%%%%%%%%%%%%%%%%%

\section{Introduction}
\label{sec:intro}

The importance of security is nowadays well recognized and mechanisms
to enforce it are being developed and adopted within
enterprises. However, this is not sufficient to ensure that security
requirements are met, as such mechanisms have to be correctly
configured and maintained at operations time. In fact, a significant
share of vulnerabilities results from security misconfiguration, as
shown by data breach reports such as~\cite{7sa10}, \cite{Ver09} and
projects such as the OWASP Top 10~\cite{Wil11}.  The reason is that
activities targeting the creation and maintenance of a secure setup,
such as patch or configuration management, are labor-intense and
error-prone.  Software vendors, for instance, issue an increasing
number of security advisories, while users, on the other hand,
struggle to understand if a given vulnerability is exploitable under
their particular conditions and requires immediate patching. As
another example, configuration best-practice provided as prose
documentation and supposingly supporting system admininistrators, is
often very broad and ambiguous.

Due to such difficulties, configuration validation is needed to gain
assurance about system security, but again, often requires manual
intervention, and thus is time-consuming and limited to samples. New
trends focus on providing standards for security automation, e.g., the
Security Content Automation Protocol (SCAP, \cite{scap}), provided by
the National Institute of Standard and Technology (NIST), whose
specifications receive a lot of attention in the scope of the
configuration baseline for IT products used in US federal
agencies~\cite{scap}.  SCAP comprises a language that allows the
specification of machine-readable security checks to facilitate the
detection of vulnerabilities caused by misconfiguration. While this
represents an important step towards the standardization and exchange
of security knowledge, SCAP focus on the granularity of hosts and
operating systems, and as such cannot be easily applied to
fine-granular and distributed system components\footnote{A system
  component hereby represents a single installation of a software
  component (or product) in a specific system, such as a given
  deployment of a Java Web Application in a Servlet container.}
independent from their environment, e.g., a Java Web Application
(JWA).  Furthermore, SCAP does not leverage standards and technologies
in the area of system and configuration management, in order to, for
instance, separate check logic and information about configuration
retrieval.

To address these limitations and make the advantages of SCAP available
to Web security experts, we propose a SCAP-based language and approach
for the declarative specification and execution of checks for sets of
fine-granular components depending on each other in a distributed
environment. Moreover we separate the check logic from the retrieval
of the configuration values for which we rely on existing system
management procedures and technologies, e.g., Configuration Management
Databases (CMDB) as defined in the IT Infrastructure Library (ITIL).
Each check is essentially a set of tests over software component
properties - such as the release and patch level - and configuration
settings that determine a system component's behavior. Though this is
not a limitation of the language, we focus on security checks, i.e.,
one of the most important usages is the detection of security
vulnerabilities.  As an example, the language allows the specification
of a check to express that the deployment descriptor of any JWA
deployed in a Servlet container supporting a Servlet specification
version of at least 3.0 must have the \texttt{http-only} flag enabled,
to prevent the access of client-side scripts to session cookies.

This paper is structured as follows. Sect.~\ref{sec:uc} introduces a
sample system based on common Open Source Software (OSS), introduces a
set of scenarios for configuration validation, and derives
requirements for a configuration validation language.
Sect.~\ref{sec:sota} presents state-of-the-art with regard to the
specification of security checks for software and configuration
vulnerabilities. Sect.~\ref{sec:lang} presents the configuration
validation language, while Sect.~\ref{sec:approach} describes our
approach.  The paper concludes with an outlook on future work in
Sect.~\ref{sec:concl}.

%%%%%%%%%%%%%%%%%%%%%%%%%%%%%%%%%%%%%%%%%%%%%%%%

\section{Use Case and Requirements}
\label{sec:uc}

This section outlines an example landscape composed of a custom
application on top of common OSS, and herewith prototypic for many
real-life systems. An overview about network topology and installed
software components is shown in Fig.~\ref{fig:app_system-diagram}. The
service provider ACME operates this landscape for its application
service ``eInvoice'', which allows customers to manage electronic
invoices, and to make them available to their business partners
through the Internet.  The application front-end for managing and
accessing invoices is implemented as a JWA. Instances of the
application, each dedicated to one customer,
 are deployed in Tomcat,
in customer-specific
context roots.
Tomcat instances run inside an internal 
\begin{wrapfloat}{figure}{c}{0pt}
	\centering \includegraphics[width=.45\linewidth]{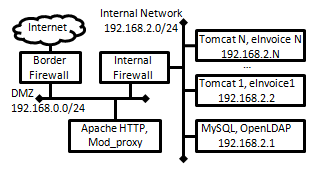}%
	\caption{ACME landscape}
	\label{fig:app_system-diagram}
\vspace{-.8cm}
\end{wrapfloat}
subnet, and are proxied by the Apache HTTP Server installed on a
physical machine connected to the DMZ. Requests for a
customer-dedicated sub-domain of acme.com are forwarded by the
reverse-proxy to the respective, customer-dedicated instance of the
JWA via the Apache JServ Protocol (AJP).

Another machine running in the internal network hosts a LDAP server
for the management of user accounts, as well as a MySQL database used
for persistency.

As the system is prototypic, so are the tasks related to configuration
management and validation. In the following, we will describe
different scenarios for configuration validation, different in terms
of periodicity, urgency (response time), validation scope, and
authorship of configuration checks.

{\bf Vulnerability Assessment (S1).}
This scenario focuses on the detection of known vulnerabilities. Upon
disclosure of a new security vulnerability of off-the-shelf applications or
software libraries, system administrators need to investigate the
susceptibility of their system. First, they need to check for the
presence of affected release and patch levels.  This can be difficult
in case of software libraries embedded into off-the-shelf applications as
their presence is often unknown. Second, they need to check whether
additional conditions for a successful exploitation are met. Such
conditions often concern specific configuration settings of the
affected software, as well as the specific usage context and system
environment.
The automation of both activities with help of machine-readable
vulnerability checks decreases time and effort required to discover a
system vulnerability, and at the same time increases the precision
with which the presence of vulnerabilities can be detected.
Precision is important as organizations are typically reluctant to
apply patches or other measures in a productive environment unless
absolutely necessary. Such checks would represent a valuable
complement to textual descriptions published by security researchers
or software vendors in vulnerability databases such as the
NVD~\cite{scap}. As an example,
CVE-2011-3190\footnote{CVE entries are maintained in vulnerability databases, e.g., NVD}
reports a vulnerability in the AJP connector implementation of several
Tomcat releases~\cite{tomcat}, which, however, only applies under certain
conditions, e.g., if certain connector classes are used, and reverse
proxy and Tomcat do not use a shared secret.
A machine check looking at the Tomcat release level and related
configuration settings could be easily provided by the application
vendor (Apache Software Foundation).
An example for a critical security bug in a software library is
CVE-2012-0392 which describes a vulnerability in Apache Struts, a
common framework to support the Model-View-Controller paradigm in
JWAs. The detection of this vulnerability is made more problematic by
the fact that end-users typically do not know if applications
installed in their environment make use of such library, and they
cannot rely on the presence of a well-established security response
process at each of their application vendors. Thus security bugs may
be dormant in libraries without the service operator being aware.

{\bf Configuration Best-Practice (S2).}  This scenario focuses on
establishing if best practices are followed.  During operations time,
system administrators need to periodically check whether the system
configurations follow best-practices, for single and distributed
system components.  Today, these are often described in prose and
evolve over time thus requiring continuous human intervention.
Examples of best-practices are the Tomcat security guide from
{OWASP}~\cite{owasp}, and the SANS recommendations for securing Java
deployment descriptors~\cite{sans}.
\begin{example}[SANS recommendation on cookie-based session handling] 
\label{ex:sans}
SANS recommends to configure the cookie-based session handling for
JWAs \linebreak[4](\verb|<cookie-config>| section of the deployment
descriptor), i.e., \emph{(i)} preventing the access to session cookies
(\verb|<http-only>| set to \verb|true|), and \emph{(ii)} transmitting
cookies securely (\verb|<secure>| set to \verb|true|).  In particular
the \verb|http-only| flag is an example of recommendation that only
applies after the release 3.0 of the Servlet specification.
\end{example}
Configuration best-practices may also cover a set of distributed
components, e.g., the how-to about Apache HTTP server
as a reverse proxy for Apache Tomcat~\cite{proxy}.
A language supporting the specification of such best-practice checks should 
support the flexible adoption to a specific
environment. A recommendation related to the session timeout, for
instance, may be refined by an organization to reflect its particular policy. 

{\bf Compliance with Configuration Policy (S3).}  This scenario
focuses on the periodic validation of landscape specific
configuration implementing the designed policy. 
Such a configuration includes a set of mandated configuration settings that an
organization expects to be active in its system.
As an example, the configuration that enforce the ACME's access control policy
embraces configuration settings of several distributed system
components, e.g., the realm definition of each Tomcat instance, as
well as the deployment descriptor of each Java application instance.
In particular the deployment descriptor has to allow the role
\verb|admin-role| to access to the URL path
\verb|/manager/*|. Moreover the realm of Tomcat has to refer to the
LDAP server located at 192.168.2.1.  This example illustrates that
configuration checks aiming to assess compliance with a given
configuration policy strongly reflect a particular system and
environment, and are therefore authored internally to the organization
rather than by externals, as in the previous scenarios.

A language for supporting the above scenarios have to fullfill the
following requirements.

\newcounter{RLcount}
\setcounter{RLcount}{0}
\begin{list}{(RL\arabic{RLcount})}
  {\usecounter{RLcount}}
\item \label{RL:1} 
  The language must support the
  definition of configuration checks for diverse software components
  (e.g., network-level firewalls or application-level access control
  systems) and diverse technologies.
\item \label{RL:2}
  The language must be expressive enough to cover new technologies or
  configuration formats without requiring extensions. This would avoid the
  need to update the language interpreter every time a new extension
  is published.
\item \label{RL:3} It must be possible to specify target components by
  defining conditions over properties such as name, release, and
  supported specification, or over the existence of relationships between
  components. This is necessary in cases where externally provided
  checks must be applied to all instances of the affected software
  components (scenarios S1 and S2).
\item \label{RL:4} Motivated by scenario S3, it must be possible to
  specify target components by referring to specific instances of a
  software component.
\item \label{RL:5} It must be possible to validate the configurations
  of different, potentially distributed system components within one
  check.
\item \label{RL:6} Checks must be uniquely identifiable, declarative,
  standardized and certifiable, to support trusted knowledge exchange
  among security tools and stakeholders, e.g., software vendors,
  experts, auditors, or operations staff. 
\item \label{RL:7} The language must support parametrization in order
  to adopt externally provided checks to a specific configuration
  policy.
\item \label{RL:8} The specification of checks must be separated from
  the collection of the involved configuration settings from a given
  managed domain.
\end{list}

%%%%%%%%%%%%%%%%%%%%%%%%%%%%%%%%%%%%%%%%%%%%%%%%

\section{State of the Art}
\label{sec:sota}

Prior art for the definition of the configuration validation language
comprises several specifications out of the Security Content
Automation Protocol (SCAP), as well as proprietary languages supported
by vulnerability and patch scanners.

SCAP~\cite{Wal11} 
is a suite of specifications that support
automated configuration, vulnerability and patch checking, as well as
security measurement. Some of the specifications are widely applied in
industry, e.g., the Common Vulnerabilities and Exposures (CVE,
\texttt{http://cve.mitre.org}), and those related to configuration
validation will be discussed with regard to above-described
requirements.
Note that several approaches assess a system's overall
security level by analyzing and reasoning about the potential
combination of individual vulnerabilities (exploits) by an
adversary~\cite{Chen2008}, \cite{Ou2005}. Though referring to SCAP
specifications, these approaches do not look into the vulnerability
specification itself, but use the language and related tools merely
for the discovery of individual vulnerabilities.

{\bf Common Platform Enumeration} (CPE,
{\texttt{http://cpe.mitre.org})} is a {XML-based} standard for the
specification of structured names for information technology systems,
software, platforms, and packages. It allows the definition of names
representing classes of platforms which can be compared in order to
establish if, e.g., two names are equal or if one of the names
represents a subset of the systems represented by the other. CPE 2.3,
the latest version, consists of four modular specifications which work
together in layers: \emph{(i)} CPE Naming providing a formal name
format, \emph{(ii)} CPE Language allowing the description of complex
platforms, \emph{(iii)} CPE Matching providing a method for checking
names against a system, and \emph{(iv)} CPE Dictionary binding text
and tests to a name.

While the specifications CPE Naming and CPE Matching allow the
definition and comparison of single software components according to
properties such as vendor or product name, the CPE Language
specification does not meet (RL3) with regard to component
relations. It supports the specification of a complex platform through
a logical condition over several CPE Names, but the semantics of their
relationship is not explicitly defined. The typical interpretation
used in many CVE entries is that a complex platform condition is met
as soon as all software components are installed on the same
machine. This interpretation, however, is in many cases not sufficient
to state that a vulnerability exists.  CVE-2003-0042, for instance, is
only exploitable if Tomcat actually uses a given JDK version, the mere
presence of both components on the same system is not sufficient. This
interpretation is even more misleading if vulnerabilities are caused
by combinations of client-side and server-side components, e.g.,
CVE-2012-0287.  A special kind of relationship is the composition of
software components, e.g., in the case of Java libraries. Today, each
vendor of an application that embeds a vulnerable library needs to
issue a dedicated CVE, as CPE insufficient to detect the use of a
given library (in an application).

{\bf Open Vulnerability Assessment Language} (OVAL,~\cite{Mit12})
defines a language for the definition of security tests detecting the
presence of vulnerabilities or configuration issues on a computer
system (machine).  It defines several XML schemas: \emph{(i)} OVAL
System Characteristics represent system configuration information that
is subject to testing, \emph{(ii)} OVAL Definitions specify conditions
for the presence of a specified machine state (vulnerability,
configuration, patch state, etc.), \emph{(iii)} OVAL Results report
the assessment result, i.e., the comparison of OVAL Definitions and
OVAL System Characteristics.

Since OVAL already fulfills some of the before-mentioned requirements,
the language proposed in Sect.~\ref{sec:lang} is to a good extent
based on OVAL concepts. According to SCAP design goals, the language
supports standardized, unambiguous, and exchangeable representations
of configuration checks (RL6) as well as variables for parametrization
(RL7).  However, a significant limitation is that OVAL checks (like
CPE) work on the granularity of machines (computer systems).  This
impacts several other requirements. With regard to (RL1), it is
difficult, sometimes impossible, to write configuration checks for
fine-granular system components independently from their software
computing environment (container), e.g., JWAs. The reason is that
generic OVAL objects from the \texttt{independent schema} (e.g.,
\texttt{textfilecontent54\_object}) are relative to the machine's file
system, which varies from one Servlet container to the other The
definition of container-specific objects (e.g.,
\texttt{spwebapplication\_object} for Microsoft Sharepoint), on the
other hand, restricts the use of checks to dedicated environments.
Requirement (RL2) is not fulfilled as OVAL requires the extension of
several schemas to address new software components. This either
requires tool vendors to constantly update the language interpreter,
or leads to a fragmented market where tools only support a subset of
the language.  We believe that the broad adoption of OVAL could be
reached more easily by the use of generic types (RL2), e.g., on the
basis of XML, herewith leveraging the fact that it is used for many
application-level configuration formats.  With regard to (RL3), (RL4)
and (RL5) it is impossible in OVAL to specify a target for checks that
look at distributed components, since the execution of a set of OVAL
definitions and their tests are meant to be executed on a single
machine.  Furthermore, OVAL does not clearly separate check logic from
the retrieval of the actual configuration values (RL8), herewith
missing to leverage industry efforts in the area of IT Service and
Application Management (ITSAM). The deployment descriptor of a JWA,
for instance, can be retrieved by several means and potentially from
different sources (the actual component, or a configuration store with
copies). The mixture of these concerns makes the work of check authors
difficult and error prone, as they cannot focus on the check logic
(e.g., the session configuration of a deployment descriptor), but also
care for the retrieval of values, e.g., the identification of a file
path depending on installation directories and environment
variables. To allow the separation of these concerns, the check
language itself must be agnostic to potential configuration sources,
the latter being cared for by administrators.

As representative vulnerability and patch scanner, we consider Nessus
({\url{http://www.tenable.com/products/nessus}}), which is a widely
adopted tool and comes with a proprietary syntax for the definition of
so-called audit checks. Organizations can either write custom checks
according to this language, or subscribe to a commercial feed to
receive compliance checks tailored for a variety of standards and
regulations, e.g., PCI
DSS~(\url{https://www.pcisecuritystandards.org}). Having comparable
expressivity, checks written in this proprietary language can be
transformed into SCAP content, which is why Nessus and similar tools
were SCAP-validated by the MITRE. SCAP and Nessus' proprietary
language also have in common that they focus on operating systems,
which makes it difficult to specify checks on a more fine-granular
level, i.e., for objects which cannot be easily identified relative to
the OS: \emph{custom items} for Windows and Unix require, for
instance, the specification of file paths which is not necessarily
possible for JWA or Web services; \emph{built-in} checks for Unix hide
the configuration source from the check author, but instead of making
the source customizable, it is hard-coded (RL9). Checks considering
distributed system components are not supported at all (RL5). Nessus
does also not allow to condition the applicability of the check on the
basis of component properties (e.g., release level) or component
relationships (RL3) but only on the basis of hard-coded keywords such
as \texttt{Unix}. As a proprietary language, processed only by Nessus,
it is not extensible by 3rd parties (RL3), nor standardized (RL4).

%%%%%%%%%%%%%%%%%%%%%%%%%

\section{Configuration Validation Language}
\label{sec:lang}
The configuration validation language allows the definition of checks
for selected software components and addresses the use cases presented
in Sect.~\ref{sec:uc}. It includes the definition of the checks as
well as of their results. This section introduces all the concepts
used within the language, and defines the extensions we carried out
over the OVAL standard. We formally define the semantics of the language without
binding to a specific syntax. Notice that in the definitions we only
consider the parts of the OVAL standard which are extended by our
language. As OVAL is XML-based, a straightforward implementation of
our formalism is an XML serialization.

\begin{figure}[!t]%
	\centering \includegraphics[width=\linewidth]{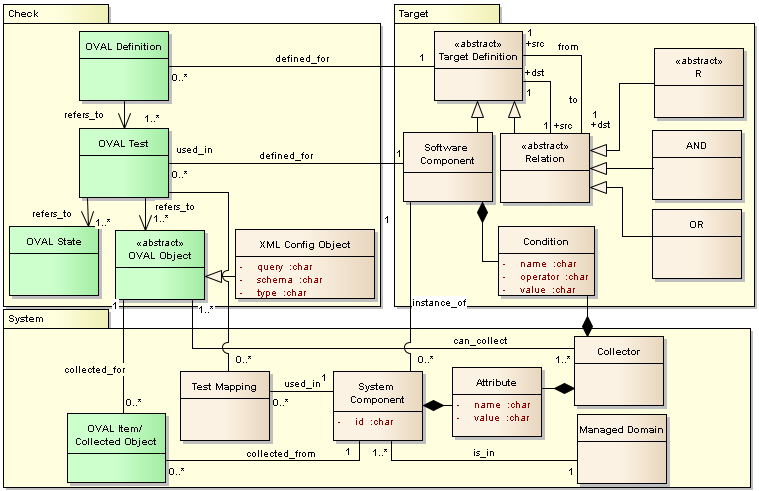}%
	\caption{Configuration validation language class diagram}
	\label{fig:lang-diagram}
\vspace{-.5cm}
\end{figure}

Fig.~\ref{fig:lang-diagram} shows the main concepts of the
configuration validation language. 
The concepts are organized into three main areas. 
The Check and Target areas concern the definition of the configuration
checks and of the affected software components, resp., the System area
contains elements corresponding to actual configurations and
components of a managed domain.

The \emph{Check} area (top left of Fig.~\ref{fig:lang-diagram})
concerns the definition of checks in the form of tests comparing an
expected and an actual value. This area relies on the OVAL
standard~\cite{Mit12}.  The concepts we borrow and extend are shown in
Fig.~\ref{fig:lang-diagram} and prefixed with ``OVAL''.
In a nutshell, a definition is characterized by an arbitrary complex boolean combination
of tests and a test defines an evaluation involving an object
(possibly containing a set of other objects) and zero or more states.
As described in Sect.~\ref{sec:sota}, the existing OVAL objects do not
fulfill requirements (RL2), and (RL8). To fulfill them, we defined a
new test, object, and state, generic enough to apply to multiple
configurations of multiple software components and independent from
the collection mechanisms. The test and state we defined are not shown
in Fig.~\ref{fig:lang-diagram} as it is the object,
\verb|XML Config Object|, that contains the major contributions. The
\verb|XML Config Object| is characterized by three attributes: the
\verb|type| denoting a type of configuration relevant for a software
component, the \verb|schema| denoting the format in which the
configurations are represented, and \verb|query| expressing how to
identify the object within the configuration. Such object also
overcomes the OVAL drawbacks about (RL1) discussed in
Sect.~\ref{sec:sota}.
\begin{example}[Object, state, and test for http-only flag]
\label{ex:flag}
The \verb|XML Config Object| can be used to specify the recommendation
described in Ex.~\ref{ex:sans}. In the excerpt below, type (line 2)
indicates that the configuration considered is a deployment descriptor
(computing environment independent), schema (line 3) refers to the
location of the schema for the deployment descriptor of J2EE web
application and the Xpath query (line 4) points to the \verb|http-only| configuration.
\lstset{basicstyle=\ttfamily\scriptsize}\lstset{numbers=left}
\begin{lstlisting}
<xmlconfiguration_object id="oval:sans.security:obj:1">
  <type>deployment descriptor</type>
  <schema>http://java.sun.com/xml/ns/j2ee</schema>
  <query>//*session-config/*cookie-config/*http-only/text()</query>
</xmlconfiguration_object>
\end{lstlisting}
By modifying only the query element, all the other recommendation of
Ex.~\ref{ex:sans} can be specified. Moreover, by modifying also
the type and schema, our object can be used for any other XML based
configuration.
The expected value for the configuration is defined through a \verb|xmlconfiguration_state|
defining \verb|true| as expected value for the
\verb|http-only| tag.
Finally, the OVAL test, \verb|xmlconfiguration_test|, contains the
object and state above which are used to evaluate the configuration.
\end{example}

\begin{definition}[OVAL Definition]
\label{def:od}
An OVAL Definition $\OD\subseteq \mathcal{T}$ is a set of OVAL
Tests.
\end{definition}

\begin{example}[OVAL Definition for SANS] 
\label{ex:od}
The OVAL definition checking for the SANS recommendations described in
Ex.~\ref{ex:sans} is a set of tests, one for each recommendation, i.e,
$\OD_{\mathit{sans}}=\{t_{http-only},t_{secure-flag}\}$.

According to OVAL, a definition is a boolean combination of tests. As
SANS requires all recommendations to be followed, all the tests
involved are characterized by an OR boolean relation in order to raise
an alarm whenever one of the recommendation is not
followed. $t_{http-only}$ (line 3) is described in
Ex.~\ref{ex:flag}. All other tests can be analogously defined.
\lstset{basicstyle=\ttfamily\scriptsize}\lstset{numbers=left}
\begin{lstlisting}
<definition id="oval:sans.security:def:1">
  <criteria operator="OR">
   <criterion test_ref="oval:sans.security:tst:1" comment="HttpOnly flag"/>
   <criterion test_ref="oval:sans.security:tst:2" comment="Secure flag"/>
  </criteria>
</definition>
\end{lstlisting}
\end{example}

The \emph{Target} area (top right of Fig.~\ref{fig:lang-diagram})
allows the definition of targets for the checks. A \emph{target
  definition} is an abstract concept representing either a software
component or a relation which can be defined over software components
or relations themselves.
A \emph{software component} is characterized by a set of conditions on
specific properties such as those in Tab.~\ref{tab:propr} (left side).
A \emph{relation} defines the relationship between software
components. We distinguish three kinds of relations. A static
relation, i.e., ``composed of'', which allows to represent the
internal structure of a software.
Run-time relations, i.e., ``deployed in'' and ``communicates with'',
which allow to define relations among software components running in a
landscape. Finally, boolean relations (AND, OR) combine either static
or dynamic relations.
Dynamic and boolean relations can be nested whereas the static
relation can only be applied to software components.  These types of
relations, combined with the possibility to nest them, allow to define
a set of software components satisfying an arbitrary complex
expression.
\begin{definition}[Software Component]
\label{def:sc}
A software component $\SC\subseteq \mathcal{C}$ is a set of
conditions. A condition $C\in\mathcal{C}$ is a tuple $C=\langle
P,\theta,V\rangle$, where
\begin{itemize}
\item $P\in\mathcal{P}$ is a property name,
\item $\theta\in\{ =, <, > , \geq , \leq\}$ is an operator,
\item $V\in \dom(P)$ is a value for the property.
\end{itemize}
\end{definition}
\begin{table}[t]
\caption{Properties description}
\label{tab:propr}
\centering \footnotesize
\begin{tabular}{|p{.13\textwidth}|p{.35\textwidth}||p{.13\textwidth}|p{.35\textwidth}|}
\hline
$\product$  & Product name, e.g., Struts&$\uncpath$ & UNC path for shared location  \\\hline
$\vendor$ & Product vendor, e.g., Apache &$\croot$ & JWA context root \\\hline
$\release$ & Product release, e.g., 2.3.1.1 &$\ipadd$ & IP address of JMX endpoint \\\hline
$\supspec$, $\reqspec$ & Supported/Required specification  &$\port$ &  Port number of JMX endpoint\\\hline
\end{tabular}
\end{table}

We define $\mathcal{R}$ as a set of relations. Examples 
are listed in Tab.~\ref{tab:rel}.
\begin{table}[!t]
\caption{Relations description}
\label{tab:rel}
\centering \footnotesize
\begin{tabular}{|l|p{.3\textwidth}||l|p{.4\textwidth}|}
\hline
$\deplin$  & deployed in, models a component installed in another &$\compof$ & composed of, represents the internal structure of applications (e.g. linked libraries)\\\hline
$\commwith$ & communicates with, represents network communication & $\inset$ & instruction set, for either compiled (x86, x64) or interpreted (Java Runtime) binaries\\\hline
\end{tabular}
\end{table}
We define $\hat{\mathcal{R}}=\mathcal{R}\times\mathbb{N}$ as the set
of numbered relations where any relation can occur an arbitrary number
of times and is uniquely identified by a natural number. In the examples we omit the natural number when no ambiguity arises. 

\begin{definition}[Target Definition]
\label{def:td}
A target definition is a tuple $\TD=\langle \SCS,\RS,\rho\rangle$ where
\begin{itemize}
\item $\SCS$ is a set of software components (cf. Def.~\ref{def:sc}),
\item $\RS\subset \hat{\mathcal{R}}$ is a set of numbered relations,
\item $\rho:\RS\rightarrow (\SCS\cup
  \RS)\times (\SCS\cup \RS)$ is a total and acyclic
  function mapping a relation into the pair of
  elements, denoted as $\rho_1$ and $\rho_2$, sharing the relation
  (either software components or relations).
\end{itemize}
A target definition $\TD=\langle \SCS,\RS,\rho\rangle$
is \emph{valid} iff $|\SCS|=1$ when $\RS=\emptyset$.
\end{definition}

\begin{example}[Software Component and Target Definition for SANS] 
\label{ex:td}
SANS applies to JWAs developed
according to one of the releases of the Servlet specification and
deployed in a web application container supporting such
specification. In particular the recommendations in
Ex.~\ref{ex:sans} refer to the release 3.0.
According to Def.~\ref{def:sc}, a software component for the web
application container can be defined
as the set containing a single condition referring to the supported
specification, $\SC_{webappcont} = \{ \langle
\supspec,\geq,\mathsf{Java\_Servlet\_3.0}\rangle \}$. As the
recommendation applies to all JWAs therein deployed, the software
component for the web application can be specified as an empty set
$\SC_{webapp}=\emptyset$.  Finally, the target definition, according
to Def.~\ref{def:td}, can be expressed as
$\TD_{sans}=\langle \SCS_{sans},\RS_{sans},\rho_{sans}\rangle$ where
$\SCS_{sans}=\{\SC_{webapp},\SC_{webappcont}\}$, $\RS_{sans}=\{\deplin\}$, and
${\rho_1}_{sans}(\deplin)=\{\SC_{webapp}\}$,
 ${\rho_2}_{sans}(\deplin)=\{\SC_{webappcont}\}$.
\end{example}

We extend the OVAL standard by referring each OVAL definition to a
target definition, i.e., to a set of related software components, and
referring each OVAL test contained in the definition to a software
component of the target definition. Thus we fulfill requirements (RL3)
and (RL5).  We name the resulting new artifact \emph{check
  definition}.  Note that this artifact is not represented by a single
class in Fig.~\ref{fig:lang-diagram} but it involves several of the
concepts therein presented and formalized above. Def.~\ref{def:od} and
\ref{def:td} provide the building blocks for the check definition.

\begin{definition}[Check Definition]
\label{def:cd}
A check definition is a tuple $\CD=\langle
\OD,\TD,\tau\rangle$ where
\begin{itemize}
\item $\OD\subseteq \mathcal{T}$ is an OVAL definition,
\item $\TD=\langle \SCS,\RS,\rho\rangle$ is a target definition,
\item $\tau: \OD \rightarrow \SCS$ is a total function
  that maps an OVAL test included in the definition $\OD$ into
  the software component to which it applies defined for the target
  definition $\TD$.
\end{itemize}
\end{definition}

\begin{example}[Check Definition for SANS] 
\label{ex:cd}
Given $\OD_{sans}$ and $\TD_{sans}$ defined in Ex.~\ref{ex:od} and
Ex.~\ref{ex:td} resp., a check definition for SANS recommendations on
cookies is $\CD_{sans}=\langle
\OD_{sans},\TD_{sans},\tau_{sans}\rangle$ where $\tau(t)=\SC_{webapp}$
for all $t\in\OD$.
\end{example}

The \emph{System} area (bottom of Fig.~\ref{fig:lang-diagram})
contains the concepts characterizing systems in a landscape and their
configurations.  A \emph{system component} represents a single
installation of a software component in a specific domain. As the
purpose is to identify its configurations, the system component is
defined as a set of attributes denoting how the configurations can be
retrieved.  The configurations required are given by the OVAL tests
which are defined for software components. To evaluate the tests, the
objects they contain have to be retrieved for each installation of the
software component, i.e., for each system component.  The tests to be
performed on system components are defined through the test
mapping. The set of attributes necessary to collect a configuration is
given by the collector (more details about how system components are
derived starting from the target definition and the collector can be
found in Sect. \ref{sec:approach}).  By allowing the separation of the
check logic from the attributes needed for the collection, our
language fulfills requirement (RL8).

\begin{definition}[Collector]
\label{def:c}
A collector is a tuple $\CL=\langle \SC_K,\PS,O_K\rangle$ where
$\SC_K$ is a set of conditions, $\PS\subseteq\mathcal{P}$ is a set of
properties, and $O_K$ is a query over OVAL objects.
\end{definition}

\begin{example}[Collector for Web Applications deployment descriptor]
\label{ex:c}
A collector for web applications deployment descriptor has to define
the set of attributes for retrieving the deployment descriptor of the
web application installed in the landscape. Several alternatives are
viable, e.g., accessing a shared file system via the Universal Naming
Convention (UNC)
or relying on the JMX interface of Tomcat. These alternatives can be
defined as two collectors, 
$\CL_{unc}=\langle \SC_{K_{webapp}},\{\uncpath\},O_{K_{webapp}}\rangle$
$\CL_{jmx}=\langle \SC_{K_{webapp}},\{\croot,\ipadd,\port\},O_{K_{webapp}}\rangle$
where
$\SC_{K_{webapp}}=\{\langle\reqspec,=,\mathsf{Java\_Servlet\_3.0}\rangle\}$
is the same for both as they apply to the same software component, and
$O_{K_{webapp}}$ is an Xpath query over the XML serialization of the
object (omitted for the sake of brevity).
\end{example}

\begin{definition}[System Component]
A system component $\SI\subseteq \mathcal{A}$ is a set of
attributes. An attribute is a tuple $A=\langle P, V\rangle$, where
$P\in\mathcal{P}$ and $V\in \dom(P)$ are properties and values, resp.
\end{definition}

\begin{example}[System Component for SANS] 
\label{ex:si}
The check definition for SANS in Ex.~\ref{ex:cd}
includes the software component $\SC_{webapp}=\emptyset$ defined in
Ex.~\ref{ex:td} which is 
referred to by an \verb|XML Config Test|. Moreover the web application
installed in the managed domain of Fig.~\ref{fig:app_system-diagram} are
characterized by the property of supporting the Servlet
specification 3.0. Thus the collector defined in Ex.~\ref{ex:c} can be
used for establishing the set of attributes of the resulting system
components.
By using $K_{unc}$, the resulting system component for
one installation of the eInvoice web application sold by ACME is
$\SI_{unc}=\{\langle \uncpath, \text{\textbackslash\textbackslash}\mathsf{192.168.2.3}\text{\textbackslash}\mathsf{ path}\text{\textbackslash}\mathsf{ to}\text{\textbackslash}\mathsf{ web.xml} \rangle\}$.
By using $K_{jmx}$, the resulting system component is 
$\SI_{jmx}=\{\langle \croot, \mathsf{/manager/*}\rangle,
\langle \ipadd, \mathsf{192.168.2.2}\rangle,\langle \port, \mathsf{8059}\rangle\}$.
\end{example}

\begin{definition}[System Test]
A system test is $\ST=\langle \SIS, \OD, \TM \rangle$
where
\begin{itemize}
\item $\SIS$ is a set of system components,
\item $\OD\subseteq\mathcal{T}$ is an OVAL definition, i.e., a set of tests,
\item $\TM\subseteq \OD\times \SIS$ is a set of test mappings defining
  which test of the definition applies to which system component.
\end{itemize}
\end{definition}

\begin{example}[System Test for SANS] \label{ex:st}
The check definition $\CD_{sans}=\langle
\OD_{sans},\TD_{sans},\tau_{sans}\rangle$ defined in Ex.~\ref{ex:cd}
originates several system tests, one for each set of software
components installed in the managed domain fulfilling the target
definition $\TD_{sans}$.  Given,
$\OD_{\mathit{sans}}=\{t_{http-only},t_{secure-flag}\}$,
$\TD_{sans}=\langle \SCS_{sans},\RS_{sans},\rho_{sans}\rangle$, and
$\tau(t)=\SC_{webapp}$, a system test defining the tests to be
performed for one possible installation of the software components is
$\ST_{sans}=\langle \SIS_{sans}, \OD_{sans}, \TM_{sans} \rangle$ 
where
$\SIS_{sans}=\{\SI_{jmx}\}$,
and $\TM_{sans}=\{(t_{http-only},\SI_{jmx}),(t_{secure-flag},\SI_{jmx})\}$.
Notice that no system component for 
$SC_{webappcont}$ is included in $\SIS_{sans}$ as no tests apply
to it.
\end{example}
The system test refers a test to specific system, thus (RL4) is met.

Finally, the \emph{OVAL Item} in Fig.~\ref{fig:lang-diagram}
represents the configuration collected from a system component for the
OVAL object defined in the OVAL test. By evaluating such items
according to the test, a boolean result for the test is
produced. Based on the test results, the boolean result of the
definition is also evaluated. Differently from OVAL, our OVAL Items
may derive from different system, however this does not affect the
evaluation algorithm defined in~\cite{Mit12}, which we rely on.
A check definition originates several system
tests, each one originating a check result.  

\begin{definition}[Check Result]
A check result is a tuple $\CR=\langle \ST, \omega\rangle$
where
\begin{itemize}
\item $\ST=\langle \SIS, \OD, \TM \rangle$ is a system test,
\item $\omega: \TM\rightarrow \{\top,\bot\}$ is a function that maps
  test mappings into its result, i.e., the boolean values true
  ($\top$) or false ($\bot$).
\end{itemize}
\end{definition}

%%%%%%%%%%%%%%%%%%%%%%%%%%%%%%%%%

\section{Approach}
\label{sec:approach}

The language presented in Sect. \ref{sec:lang} separates the
checks' logic from the systems to which they apply. In this
section we establish the link between these two aspects, thereby describing how 
the checks can be instantiated and executed in a concrete landscape. 

The overall approach is outlined in Fig.~\ref{fig:approach}. 
External and
internal authors (from the perspective of an organization)
can define, independently from the landscape, checks $\CD$
(Def.~\ref{def:cd}) for known vulnerabilities affecting software
components (cf. (S1)), and for best practices of single or multiple
software components sharing relations (cf. (S2)).
An additional input
is the
set of collector definitions $\mathcal{K}$, that has to be provided by
system administrators as creates the link between the software
components used in the checks and the attributes of system components
which allow the collection of the configurations. 
The \verb|TD Evaluator| module
has in input the above artifacts and is responsible for producing all
the system tests $\ST$ defining which test has to be executed on
which system component.
To produce the System Test
artifact, the \verb|TD Evaluator| relies on a \verb|Data Source|, an
authoritative source of information about the software components
installed in a \mandom. 
We assume a single \verb|Data Source| to provide
information about several aspects of the \mandom, ranging from the
properties of installed software (e.g. product names and vendors), or
the internal structure of applications (e.g. linked libraries), up to
architectural details on the deployment or the network interaction among
different pieces of software. Since such information is often
scattered over several repositories within an organization (e.g., CMDBs),
the \verb|Data Source| is a federated set of views over these
repositories, which constitute the interface to our language.
\begin{wrapfloat}{figure}{c}{0pt}
\includegraphics[scale=0.4]{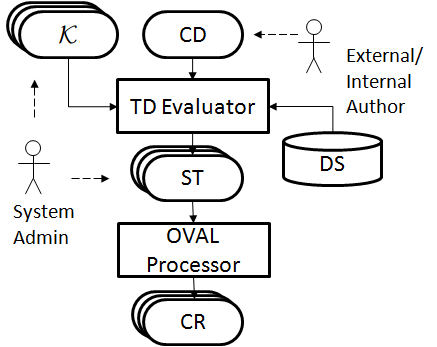}
	\caption{Detection of vulnerabilities approach.}
	\label{fig:approach}
\vspace{-.6cm}
\end{wrapfloat}
Although strong, this assumption is not unrealistic. Indeed, several theoretical formulations
of this problem are tackled in literature on data integration~\cite{Ullman97}\cite{Lenzerini2002}.
Furthermore the increasing adoption of standards such as DMTF's CMDBf~\cite{cmdbf} demonstrates the
practical feasibility of configuration data federation.

The system test can also be manually provided by system administrators
in case of checks for selected system components (cf. scenario
(S3)). System tests are then processed by the \verb|OVAL Processor|
module that interprets the OVAL content and collects the objects
defined for each system component within $\ST$. The configurations
collected from distributed systems are then evaluated and 
check results $\CR$ are produced, highlighting
existing misconfiguration issues (if any).

A key step of the approach is the generation of the system tests based
on the data source.
In the following we formally define the interpretation of
\targetdef s  w.r.t. a \Ds, which provides information about the properties of \softcomp s deployed within a \mandom. 
We then describe how this leads to the generation of system tests. 

Informally a \Ds\ can be seen as a particular instantiation of
\softcomp\ \propts\ (cf. Def.~\ref{def:sc}) and \targetdef\ \relt s
(cf. Def.~\ref{def:td}) for a \mandom.  Let $\mathcal{I}$ be the
domain of instances of software components, namely \sysinst s,
containing one unique symbol for each software component installed in
a given \mandom. The \Ds\ then maps every \sysinst\ to the actual
values of its properties and links it to the other \sysinst s it is
related to.

\begin{definition}[Data Source]
A \Ds\ is the pair of sets $\ds = \langle \Pi, \Gamma \rangle$. $\Pi$ contains a partial function $\pi_P : \mathcal{I} \rightarrow \dom(P)$ for each \propt\ $P \in \mathcal{P}$, while $\Gamma$ includes a relation $\gamma_R \subseteq \mathcal{I} \times \mathcal{I}$ for each symbol $R \in \mathcal{R}$. 
\end{definition}

\begin{example}[Data Source]
\label{ex:dsi}
Figure~\ref{fig:dsinst} depicts a tabular representation of the \Ds\ $DS_1$ for the example landscape of Fig.~\ref{fig:app_system-diagram}. Due to space limitations, only a subset of the \propts\ listed in Tab.~\ref{tab:propr} and \relt s of Tab.~\ref{tab:rel} are considered.
\begin{figure}[ht]
\centering
\horizfig[$\pi_{\vendor}$]{.16\textwidth}{
\begin{tabular}{||c|c||}\footnotesize
$i$ & $\pi_{\vendor}(i)$\\
\hline
$a$ & \textsf{Apache}\\
$l$ & \textsf{OpenLDAP}\\
$t_1$ & \textsf{Apache}\\
$t_2$ & \textsf{Apache}\\
$w_a$ & \textsf{ACME}\\
$w_b$ & \textsf{ACME}\\
$w_c$ & \textsf{ACME}\\
\end{tabular}}
\hfill
\horizfig[$\pi_{\release}$]{.15\textwidth}{
\begin{tabular}{||c|c||}\footnotesize
 $i$ & $\pi_{\release}(i)$\\
\hline
$a$ & \textsf{2.2}\\
$l$ & \textsf{2.4.30}\\
$t_1$ & \textsf{7.0.18}\\
$t_2$ & \textsf{7.0.18}\\
$w_a$ & \textsf{1.0}\\
$w_b$ & \textsf{1.0}\\
$w_c$ & \textsf{1.0}\\
\end{tabular}}
\hfill
\horizfig[$\pi_{\product}$]{.22\textwidth}{
\begin{tabular}{||c|c||}\footnotesize
$i$ & $\pi_{\product}(i)$\\
\hline
$a$ & \textsf{Apache HTTPd}\\
$l$ & \textsf{OpenLDAP}\\
$t_1$ & \textsf{Tomcat}\\
$t_2$ & \textsf{Tomcat}\\
$w_a$ & \textsf{Web eInvoice}\\
$w_b$ & \textsf{Web eInvoice}\\
$w_c$ & \textsf{Web eInvoice}\\  
\end{tabular}}
\hfill
\horizfig[$\pi_{\supspec}$]{.2\textwidth}{
\begin{tabular}{||c|c||}\footnotesize
$i$ & $\pi_{\supspec}(i)$\\
\hline
$t_1$ & \textsf{Java\_Servlet\_2.5}\\
$t_1$ & \textsf{Java\_Servlet\_3.0}\\
$t_2$ & \textsf{Java\_Servlet\_2.5}\\
$t_2$ & \textsf{Java\_Servlet\_3.0}\
\end{tabular}}
\hfill
\horizfig[$\gamma_{\deplin}$]{.12\textwidth}{
\begin{tabular}{||c|c||}\footnotesize
$i_1$ & $i_2$\\
\hline
$w_a$ & $t_1$\\
$w_b$ & $t_2$\\
$w_c$ & $t_2$\\
\end{tabular}}
\caption{Example of \Ds\ instance.\label{fig:dsinst}}
\vspace{-.7cm}
\end{figure}
\end{example}

A \softcomp\ can be seen as a simple conjunctive query ranging over properties of software deployed within a \mandom. The \Ds\ provides the necessary views on the \mandom\ to answer such a query. The answer consists of the set of \sysinst s matching to all the conditions within the \softcomp. If it contains no conditions, the answer is the entire domain of \sysinst s $\mathcal{I}$. This evaluation is performed by the \Ds\ interpretation of \softcomp s, given by the mapping $\scsem{\cdot} : \mathcal{SC} \rightarrow 2^{\mathcal{I}}$:
\begin{align}
\scsem{\emptyset} & = \mathcal{I}\\ \nonumber
\scsem{\langle P,\theta,v \rangle \cup SC} & = \left\{ i \in \mathcal{I} \ | \ \pi_P(i) \ \theta \ v \right\} \cap \scsem{SC}.
\end{align}
A target definition $\TD=\langle \SCS,\RS,\tau\rangle$ is instead a more complex selection predicate
(cf. Def.~\ref{def:td}) and there can be several sets of software
component identifiers which satisfy it. The
interpretation of $\TD$ over a data source $DS$, $\llbracket \TD
\rrbracket_{DS}$, provides all such sets. This is done
by relying on two interpretation functions, one providing the sets of
software component identifiers, and one providing a function that maps
each software component identifier to the corresponding software
component.

The interpretation function $\rsem{\cdot} : (\mathcal{SC} \cup
\mathcal{\hat{R}}) \rightarrow 2^{2^{\mathcal{I}}}$ associates every
$\SC \in \mathcal{SC}$ and $R \in \mathcal{\hat{R}}$
to a powerset of \sysinst s, as defined in  \eqref{eq:scsem} and \eqref{eq:rsem}, respectively.
Notice that this function depends both on the data source $DS$ and 
the function $\rho$ that carries the structure of target definition expressions.
\begingroup
\setlength{\belowdisplayskip}{0pt} 
\begin{align}\label{eq:scsem}
\rsem{SC} & = \big\{ \scsem{SC} \big\} \\ \label{eq:rsem}
\rsem{R} & = \left\{\begin{array}{l l}
	\rsem{\rho_1(R)} \times \rsem{\rho_2(R)} & \text{if } R=\wedge \\
	\rsem{\rho_1(R)} \cup \rsem{\rho_2(x)} & \text{if } R=\vee \\
	\big \{ \{ v_1,\ldots,v_n,w_1,\ldots,w_m \} \ | \ \{ v_1,\ldots,v_n \} \in \rsem{\rho_1(R)},\\
		 \{ w_1,\ldots,w_m \} \in  \rsem{\rho_2(R)}, \langle v_{1 \le i \le n},w_{1 \le j \le m} \rangle \in \gamma_R \big \} & \text{otherwise}
\end{array}\right.
\end{align}
\endgroup

Similarly, the interpretation function $\sigsem{\cdot} : (\mathcal{SC} \cup
\mathcal{\hat{R}}) \rightarrow \left(\mathcal{I} \rightarrow
\mathcal{SC} \right)$ maps every $\SC \in \mathcal{SC}$ and $R \in
\mathcal{\hat{R}}$ to a function $\sigma$ associating each
\sysinst\ to the corresponding \softcomp, according to  \eqref{eq:sigscsem} and \eqref{eq:sigrsem}.
\begingroup
\setlength\abovedisplayskip{3pt}
\begin{eqnarray}\label{eq:sigscsem}
\sigsem{SC} & = & \sigma, \text{ where } \sigma(i) = \SC, \forall i \in \scsem{SC}\\ \nonumber
\sigsem{R} & = & \sigma, \text{ where } \sigma(i) = \sigsem{\rho_1(R)}(i), \forall i \in \dom(\sigsem{\rho_1(R)}) \\ \label{eq:sigrsem}
&& \quad \text{ and } \sigma(j) = \sigsem{\rho_2(R)}(j), \forall j \in \dom(\sigsem{\rho_2(R)}).
\end{eqnarray}
\endgroup

Finally, the evaluation function for a valid \targetdef\ $\TD =
\langle \SCS, \RS, \rho \rangle$ over the \Ds\ $DS$,
$\llbracket \cdot \rrbracket_{DS} : \mathcal{TD} \rightarrow
\left(2^{2^{\mathcal{I}}} \times \left(\mathcal{I} \rightarrow
\mathcal{SC} \right) \right)$, associates a $\TD$ to the pair
$\langle I^{*}, \sigma \rangle$, where $I^{*}$ is a powerset of
\sysinst s and $\sigma$ a function mapping every $i \in I \in I^{*}$
to a $\SC \in \SCS$. As expressed in \eqref{eq:tdsem}, the
definition of $\llbracket \cdot \rrbracket_{DS}$ relies on the
aforementioned recursive interpretation functions of all the elements
within the \targetdef\ expression, starting, in the general case, from
the only relation
$R_0$ which never appears in the $\rho$ co-domain. In case $\RS =
\emptyset$, we know from Def.~\ref{def:td} that $\exists ! \ \SC_0 \in
\SCS$ and therefore $\SC_0$ is the only element being interpreted.

\begin{equation}\label{eq:tdsem}
\llbracket \TD \rrbracket_{DS} = 
\left\{\begin{array}{l l}
		\langle\rsem{R_0},\sigsem{R_0}\rangle, \text{with } \{R_0\} = \dom(\rho) \setminus \cod(\rho) & \text{ if } \RS \neq \emptyset\\
		\langle\rsem{\SC_0},\sigsem{\SC_0}\rangle, \text{with } \{\SC_0\}=\SCS  & \text{otherwise}.\\
\end{array} \right.
\end{equation}

\begin{example}\label{ex:tdsem}
We hereby compute the interpretation of the \targetdef\ $\TD_{sans}$, introduced in Ex.~\ref{ex:td}, w.r.t. the \Ds\ $DS_1$, shown in Ex.~\ref{ex:dsi}.

First, we recognize (Eq.~\eqref{eq:tdsem}) that $\llbracket TD_{sans} \rrbracket_{DS_1} = \langle I^{*}_{sans}, \sigma_{sans} \rangle = \langle \rsemp{\deplin}{DS_1,\rho}, \sigsemp{\deplin}{DS_1,\rho} \rangle$, since $\deplin \in \dom(\rho) \setminus \cod(\rho)$.

In order to obtain $\rsemp{\deplin}{DS_1,\rho}$, according to \eqref{eq:rsem}, we now need to compute the two following terms:
\begin{enumerate}[(i)]
\item $\rsemp{\rho_1(\deplin)}{DS_1,\rho} = \{ \scsemp{\SC_{webapp}}{DS_1,\rho} \} = \{ \scsemp{\emptyset}{DS_1} \} = \{ \mathcal{I} \}$;
\item $\rsemp{\rho_2(\deplin)}{DS_1,\rho} = \{ \scsemp{\SC_{webappcont}}{DS_1,\rho} \} =$\\
$=\scsemp{\{ \langle \supspec,\geq,\mathsf{Java\_Servlet\_3.0}\rangle \}}{DS_1,\rho} =$\\
$=\{ \{i \in \mathcal{I} \ | \ \pi_{\supspec}(i) \geq \mathsf{Java\_Servlet\_3.0}\} \} = \{ \{ t_1, t_2 \} \}$.
\end{enumerate}

We then have $I^{*}_{sans} = \rsemp{\deplin}{DS_1,\rho} = \bigl\{ \{ v, w \} \ | \  v \in \mathcal{I}, w \in \{ t_1, t_2 \}, \langle v, w \rangle \in \{ \langle w_a, t_1 \rangle, \langle w_b, t_2 \rangle, \langle w_c, t_2 \rangle \} \bigr\} = \bigl\{ \{ w_a, t_1 \}, \{ w_b, t_2 \}, \{ w_c, t_2 \} \bigr\}$.

Analogously, by applying \eqref{eq:sigrsem}, we obtain $\sigma_{sans} = \sigsemp{\deplin}{DS_1,\rho} = \{ w_a : \SC_{webapp}, w_b : \SC_{webapp}, w_c : \SC_{webapp}, t_1 : \SC_{webappcont}, t_2 : \SC_{webappcont} \}$.
\end{example}

As last step,
the \verb|TD Evaluator| needs to identify one or more system tests,
mapping each OVAL test to the \syscomp\
carrying the information about how to collect the object.

A check definition $\CD = \langle \OD, \TD, \tau \rangle$ is defined
for the \targetdef\ $\TD$, being interpreted over a
\Ds\ resulting in a pair $\llbracket \TD \rrbracket_{DS} =
\langle I^{*}, \sigma \rangle$. Every $I \in I^{*}$ is a set of
\sysinst s satisfying the $\TD$ expression. Therefore one system test
has to be created for every such set $I$.

When the \verb|TD Evaluator| processes a check definition, it must
identify a \emph{matching collector} $\CL$, among the set
$\mathcal{K}$ of all the ones defined for a given \mandom. This has to
be done for every \sysinst\ $i \in I$, and provides the set of
properties $PS$ necessary to collect the to-be-checked configurations
for specific OVAL Objects from $i$. For this reason, every $\CL \in
\mathcal{K}$ (cf. Def.~\ref{def:c}) references a \softcomp\ $SC_K$ and
contains a Xpath query $O_\CL$, matching to the XML serialization of
the OVAL Objects it applies to. We write $t \models O_\CL$ whenever
the XML serialization of all the OVAL Objects referenced within an
OVAL Test $t$ satisfy the Xpath query $O_\CL$.

Given a collector property set $PS$ and a \sysinst\ $i$, Eq.~\eqref{eq:pssem} defines how to retrieve the corresponding \syscomp\ from a \Ds\ $DS$, through the interpretation function $\pssemp{\cdot}{DS}(i) : 2^{\mathcal{P}} \rightarrow \mathcal{SI}$.
\begingroup
\begin{equation}\label{eq:pssem}
\pssemp{\PS}{DS}(i) = \pssemp{\{P_1, \ldots, P_n\}}{DS,i} = \{\langle P_1,\pi_{P_1}(i) \rangle, \ldots, \langle P_n, \pi_{P_n}(i) \rangle\}.
\end{equation}
\endgroup

The conditions required to determine whether a collector matches to a \sysinst\ are now formalized by the following definition.

\begin{definition}[Matching Collector]\label{def:mcoll}
For a $\CD = \langle \OD, \TD, \tau \rangle$, where $\TD = \langle \SCS, \RS, \rho \rangle$, let $\llbracket \TD \rrbracket_{\mathcal{DS}} = \langle I^{*}, \sigma \rangle$ be an interpretation of $\TD$ over $DS$ and \linebreak[4]$\tau^{-1} : \SCS \rightarrow 2^{\OD}$ be the inverse of $\tau$, mapping every $\SC$ to the set $\{t~\in~\OD~|\ \linebreak[4]\tau(t)=\SC \}$.
We then say that~$\CL=\langle~\SC_{\CL},~\PS,~O_{\CL}\rangle$~matches~to~$i~\in~I~\in~I^{*}$,~iff
\begin{equation*}
i \in \scsem{\SC_\CL} \text{ and } P \in \PS \Rightarrow \exists \langle P,\cdot \rangle \in \pssemp{\PS}{DS}(i) \text{ and } t \in \tau^{-1}(\sigma(i)) \Rightarrow t \models O_\CL.
\end{equation*}
\end{definition}

Given the interpretation $\llbracket \TD \rrbracket_{DS} = \langle I^{*}, \sigma \rangle$ of a \targetdef\ within a check definition $\CD = \langle \OD, \TD, \tau \rangle$, we are now in a position to associate each $I \in I^{*}$ to a system test $\ST_I = \langle \SIS_I, \OD, \TM_I \rangle$, constructed as follows.
(i) $\OD$ is the same OVAL Definition contained in $\CD$.
(ii) Every element $\SI \in \SIS_I$ is a \syscomp, i.e. a collection of attributes associated to properties of the \sysinst\ which allows to collect configuration information from it. For every $i \in I$ we first need to find a matching collector $K$ carrying such set of properties $\PS$, and we then retrieve the \syscomp\ $\SI$, i.e. the attributes corresponding to the properties in $\PS$, from the \Ds\ $DS$. (iii) $\TM_I$ maps every test $t \in \OD$ to a \syscomp\ $\SI \in \SIS_I$.

Eq.~\eqref{eq:ts} finally specifies how the system test's components $SIS_I$ and $TM_I$, informally described above, are built by the \verb|TD Evaluator|.
\begingroup
\begin{equation}\label{eq:ts}
\begin{split}
\forall i \in I \text{ if } \exists \CL \in \mathcal{K} \text{ s.t. } \CL \text{ matches to } i, \text{ then }\\
\pssemp{\PS}{DS}(i) \in \SIS_I \text{ and } \langle t,\pssemp{\PS}{DS,i} \rangle \in \TM_I \ \forall t \in \tau^{-1}(\sigma(i)).
\end{split}
\end{equation}
\endgroup

\begin{example}
Let us consider the check definition $\CD_{sans}=\langle \OD_{sans},\TD_{sans},\tau_{sans}\rangle$, introduced in Ex.~\ref{ex:cd}, and the data source interpretation of its target definition $\llbracket TD_{sans} \rrbracket_{DS_1} = \langle I^{*}_{sans}, \sigma_{sans} \rangle$, which has been derived in Ex.~\ref{ex:tdsem}.
Three sets of \sysinst s satisfy the target definition, namely $I^{*}_{sans} = \bigl\{ \{w_a,t_1\}, \{w_b,t_2\}, \{w_c,t_2\} \bigr\} = \bigl\{ I_a, I_b, I_c \bigr\}$, hence three system tests will be created. Among those, we shall only discuss, for brevity, the system tests $ST_{I_a}$ and $ST_{I_b}$, related to $I_a$ and $I_b$ resp.

For the sake of this example we extend the data source $DS_1=\langle \Pi_1, \Gamma_1 \rangle$ such that it includes the properties required by the collectors (cf. Ex.~\ref{ex:c}). Let such an extended data source be $DS_1'=\langle \Pi_1 \cup \{\pi_{\croot}, \pi_{\ipadd}, \pi_{\port}, \pi_{\uncpath},\},\Gamma \rangle$, where: $\pi_{\croot}(w_a) = \mathsf{/manager/*}$, $\pi_{\ipadd}(w_a) = \mathsf{192.168.2.2}$, $\pi_{\port}(w_a) = \mathsf{8059}$, and $\pi_{\uncpath}(w_b) = \text{\textbackslash\textbackslash}\mathsf{192.168.2.3}\text{\textbackslash}\mathsf{ path}\text{\textbackslash}\mathsf{ to}\text{\textbackslash}\mathsf{ web.xml}$.

According to Def.~\ref{def:mcoll} the collector $K_{jmx}$ matches to the \sysinst\ $w_a$ (and not to $w_b$), as (i) $w_a \in \scsem{\SC_{K_{webapp}}}$, (ii) $\pi_{\croot}(w_a)$, $\pi_{\port}(w_a)$, $\pi_{\port}(w_a)$ are all defined in $DS$ (while this is not the case for $w_b$), and (iii) both $t_{http-only} \models O_{K_{webapp}}$ and $t_{secure-flag} \models O_{K_{webapp}}$ hold.
From analogous reasoning it follows that $K_{unc}$ matches to $w_b$ (and not to $w_a$).

By applying \eqref{eq:ts} we finally derive that $ST_{I_a} = \langle \{\SI_{jmx}\}, OD_{sans}, \{(t_{http-only},\SI_{jmx}),(t_{secure-flag},\SI_{jmx})\} \rangle = ST_{sans}$, as anticipated in Ex.~\ref{ex:si} and~\ref{ex:st}. Analogously, we obtain $ST_{I_b} = \langle \{\SI_{unc}\}, OD_{sans}, \{(t_{http-only},\SI_{unc}),(t_{secure-flag},\SI_{unc})\} \rangle$.
\end{example}

%%%%%%%%%%%%%%%%%%%%%%%%%%%%%%%%%%%%%%%%%%%%%%%%

\section{Conclusion and Future Work}
\label{sec:concl}

This paper presents a formal approach to
specify and execute declarative and unambiguous checks able to detect vulnerabilities
caused by system misconfiguration.  This paper extends the state of
the art on configuration validation as security checks can be
specified for fine-granular components in a distributed environment
and separate the check logic from the configuration retrieval.

A proof of concept has been developed to explore the feasibility of
our approach at the example of OWASP and SANS recommendations for JWA,
using a CMDB as data source for resolving target definitions, and JMX
for the collection of configuration settings.  In future work, we will
evaluate the prototype in near-world environments that comprise a
greater numbers of system components.  Furthermore, we plan to
generate security checks and checklists in an automated fashion to
facilitate scenario (S3), where checks are used for gaining assurance
about compliance with system-specific configuration policies. This
would allow to gain assurance without the need to manually author
check on a low technical level.  Lastly, we intent to investigate the
usage in cloud scenarios, were cloud providers could use and offer a
corresponding tool for ensuring the security of consumer-managed
resources.

%%%%%%%%%%%%%%%%%%%%%%%%%%%%%%%%%%%%%%%%%%%%%%%%


\begin{thebibliography}{10}

\bibitem{7sa10}
7Safe, the University~of Bedfordshire:
\newblock Uk security breach investigations report 2010.
\newblock \url{http://www.7safe.com/breach_report/Breach_report_2010.pdf}
  (2010)

\bibitem{Ver09}
Verizon:
\newblock 2009 data breach investigations report.
\newblock Verizon,
  \url{http://www.7safe.com/breach_report/Breach_report_2010.pdf} (2009)

\bibitem{Wil11}
Williams, J., Wichers, D.:
\newblock Top 10 most critical web application security risks.
\newblock OWASP, \url{https://www.owasp.org/index.php/Top_10_2010-A6} (2010)

\bibitem{scap}

  \url{http://{scap,usgcb,nvd}.nist.gov}

\bibitem{tomcat}

$\mathtt{  http://tomcat.apache.org/security-6.html\#Fixed\_in\_Apache\_Tomcat\_6.0.35}$

\bibitem{owasp}

  \url{https://www.owasp.org/index.php/Securing_tomcat}

\bibitem{sans}

  \url{http://software-security.sans.org/blog/2010/08/11/security-misconfigurations-java-webxml-files}

\bibitem{proxy}

  \url{http://tomcat.apache.org/connectors-doc/generic_howto/proxy.html}

\bibitem{Chen2008}
Chen, X., Zheng, Q., Guan, X.:
\newblock An OVAL-based active vulnerability assessment system for enterprise
  computer networks.
\newblock ISF (2008)  573--588

\bibitem{Ou2005}
Ou, X., Govindavajhala, S., Appel, A.W.:
\newblock MulVal: a logic-based network security analyzer.
\newblock In: USENIX Security Symposium. (2005)

\bibitem{Wal11}
Waltermire, D., Quinn, S., Scarfone, K.:
\newblock The technical specification for the Security Content Automation
  Protocol (SCAP): SCAP version 1.1.
\newblock NIST,
  \url{http://csrc.nist.gov/publications/nistpubs/800-126-rev1/SP800-126r1.pdf}
  (2011)

\bibitem{Mit12}
J.~Baker, M.~Hansbury, D.H.:
\newblock The OVAL language specification (version 5.10.1).
\newblock MITRE Corporation,
  \url{http://oval.mitre.org/language/version5.10.1/OVAL_Language_Specification_01-20-2012.pdf}
  (2012)

\bibitem{Ullman97}
Ullman, J.D.:
\newblock Information integration using logical views, ICDT, (1997) 19--40

\bibitem{Lenzerini2002}
Lenzerini, M.:
\newblock Data integration: a theoretical perspective.
\newblock PODS (2002)  233--246

\bibitem{cmdbf}
DMTF Distributed Management Task Force:
\newblock Configuration Management Database (CMDB) Federation Specification.
\newblock DMTF Technical Report DSP0252. (2010)

\end{thebibliography}
\end{document}